\documentclass[letter]{jpsj2}

\title{Markovian Approximation for the Nos\'e--Hoover method and H-theorem}

\author{Hiroshi \textsc{WATANABE}\thanks{E-mail address: hwatanabe@is.nagoya-u.ac.jp}}

\inst{
Department of Complex Systems Science,
Graduate School of Information Science,
Nagoya University
}

\abst{
A Langevin equation with state-dependent random force is considered.
When the Helmholtz free energy is
a nonincreasing function of time (the H-theorem),
a generalized Einstein relation is obtained.
A stochastic process of the Nos\'e--Hoover method is discussed
on the basis of the Markovian approximation.
It is found that the generalized Einstein relation holds for 
the Fokker--Planck equation associated with the stochastic Nos\'e--Hoover equation.
The present result indicates that the Nos\'e--Hoover dynamics
coarse-grained with time satisfies the H-theorem
and therefore works as a heat bath.
}

\kword{Nos\'e--Hoover method, Langevin equation, H-theorem, heat bath}
\newcommand{\diff}{\mathrm d}

\begin{document}
\maketitle

The purpose of the present study is to investigate the
entropy production of the system,
the temperature of which controlled by the Nos\'e--Hoover method.
The equations of motion of the Nos\'e--Hoover method
contain friction terms with memory effects.~\cite{Nose1984_MolPhys,Nose1984_JChemPhys,Hoover1985,Nose1991}.
These equations are time-reversible and the distribution function
of the system does not change with time, providing that the system is ergodic.
A system driven by the Nos\'e--Hoover method, therefore,
does not exhibit relaxation from the microscopic viewpoint,
{\it i.e.}, there is no entropy production.
However, if one observes macroscopic quantities, such as the temperature of the system,
it exhibits relaxation to the values in the equilibrium state.
This implies that the thermostated system is time-irreversible
and exhibits entropy production from the macroscopic viewpoint.

This phenomenon, time-reversible from the microscopic viewpoint 
and time-irreversible from the macroscopic viewpoint,
has been considered in the theory of generalized Brownian motions~\cite{Zwanzig1975}.
Mori derived the so-called Mori equation from Hamilton's equation
by the projection-operator method~\cite{Mori1965}.
The Mori equation describes the time evolution of macroscopic variables,
which are the functions of microscopic variables. While
the Mori equation exhibits time reversibility,
the generalized Langevin equation derived from the Mori equation
is time-irreversible because of the Markovian approximation.
Zwanzig derived the explicit and exact form of the Langevin equation from the microscopic 
dynamics of the system interacting with many harmonic oscillators~\cite{Zwanzig1973}.
This formulation clearly shows how the system loses its time reversibility
with the Markovian approximation.

The above studies consider the stochastic process of the macroscopic 
variables driven by the deterministic dynamics of the microscopic variables.
Thus, a similar analysis may be possible for the system with a
thermostat. The projection, \textit{i.e.}, the elimination of the degrees of freedom,
and Markovian approximation lead to the stochastic process of the
thermostated system, which enables us to study the entropy production
of the system with the Nos\'e--Hoover method or other thermostats.

In the present article, we first consider the generalized Langevin equation
with state-dependent random force. We derive the extended Einstein relation
by considering the monotonic decay of the free energy, {\it i.e.}, the H-theorem.
Then, we discuss the stochastic process of the system with the Nos\'e--Hoover method
by the Markovian approximation, and we show that the equation of motion with coarse graining 
is a special case of the generalized Langevin equation with state-dependent random force.
A comparison with the traditional Brownian motion theory is also given.

We start from the following equations of motion,
\begin{equation}
\left\{
\begin{array}{ccl}
\dot{p} &=& - \displaystyle \frac{\partial H}{\partial q} + \hat{r}, \\
\dot{q} &=& \displaystyle \frac{\partial H}{\partial p},
\end{array}
\right. \label{eq_motion}
\end{equation}
with a Hamiltonian $H(p,q)$ and a thermostating term $\hat{r}$.
Throughout this manuscript, stochastic variables are denoted by a hat,
and we consider the system with one-degree of freedom for simplicity.
For the system to exhibit a stable and steady state, the thermostating 
term should contain two parts, namely, a friction term and a diffusion term.
Here, we take the form of the term $\hat{r}$ to be
\begin{equation}
\hat{r}(p,t) = h(p) + g(p) \hat{R}(t),
\end{equation}
where $h(p)$ denotes the friction term and $g(p) \hat{R}(t)$ denotes the
state-dependent diffusion term with the noise $\hat{R}$.
We assume that the noise $\hat{R}$ is Gaussian,
the autocorrelation function of which is given by
\begin{equation}
{\mathrm E}( \hat{R}(t) \hat{R}(t')) = 2 D \delta (t-t'), \label{eq_noise}
\end{equation}
where ${\mathrm E}(\hat{X})$ denotes the expected value of the random variable $\hat{X}$ and
$D (>0)$ denotes the amplitude of the noise.
The Fokker-Planck equation corresponding to the equations of motion (\ref{eq_motion}) is
\begin{equation}
\frac{\partial f}{\partial t} = - \left\{H,f\right\} -\frac{\partial J_p}{\partial p}, \label{eq_fp_full}
\end{equation}
where $f$ denotes the distribution function,
curly brackets denote Poisson brackets and $J_p$ is the 
probability flux corresponding to the thermostating term.
The flux $J_p$ is defined as
\begin{equation}
J_p \equiv {\mathrm E}( \hat{r} f ).
\end{equation}
Since the noise $\hat{R}$ is assumed to be Gaussian, the flux is expressed as
\begin{eqnarray}
J_p &=& \left( h - D g \frac{\partial }{\partial p}g \right)f, \\
&=&  - D g^2 \left[ \left( - \frac{h}{D g^2} + \frac{g'}{g} \right)f + \frac{\partial f}{\partial p} \right], \label{eq_Jp}
\end{eqnarray}
with $g' \equiv \diff g/ \diff p$.

The Helmholtz free energy of this system is defined as
\begin{equation}
F(t) \equiv \int f \left(H + \beta^{-1} \ln f \right) \diff p \diff q,
\end{equation}
with the inverse temperature $\beta$.
If the system is isolated,
the H-theorem must hold, {\it i.e.}, the free energy should be a nonincreasing
function of time as
\begin{equation}
\frac{\diff F}{\diff t} \leq 0.
\end{equation}
The derivative of the free energy with respect to time is
\begin{eqnarray}
\frac{\diff F}{\diff t} &=& \int \frac{\partial f}{\partial t} \left(H + \beta^{-1} \ln f + 1 \right) \diff p \diff q,\\
&=& - \int \frac{\partial J_p}{\partial p} \left(H + \beta^{-1} \ln f \right) \diff p \diff q,\\
&=& \beta^{-1} \int \frac{ J_p}{f} \left( \beta \frac{\partial H}{\partial p} f + \frac{\partial f}{\partial p} \right) \diff p \diff q. \label{eq_int}
\end{eqnarray}
Here, we use the conservation of probability
$$ \int \frac{\partial f}{ \partial t} \diff p \diff q= 0,$$
and the integration by parts is applied with respect to $p$.
For the H-theorem to hold, the integrand of Eq.~(\ref{eq_int}) must always be
nonpositive as
\begin{equation}
\frac{1}{f} J_p \left( \beta \frac{\partial H}{\partial p}f +  \frac{\partial f}{\partial p}\right) \leq 0. \label{eq_negative}
\end{equation}
Comparing Eqs.~(\ref{eq_Jp}) and (\ref{eq_negative}), we obtain
\begin{eqnarray}
- \frac{h}{D g^2} + \frac{g'}{g} = \beta \frac{\partial H}{\partial p},
\end{eqnarray}
or equivalently, 
\begin{eqnarray}
h = - D \beta g^2 \frac{\partial H}{\partial p} + D g g' \label{eq_h}.
\end{eqnarray}
The above equation describes the relationship between the 
friction term and the diffusion term. This includes the
Einstein relation as the special case when $g$ is constant.
Equation~(\ref{eq_h}) is, therefore, one of the expressions of the 
fluctuation-dissipation theorem.
Inserting the obtained expressions for $h$ into the equation of motion~(\ref{eq_motion}),
we obtain the generalized Langevin equation,
\begin{equation}
\left\{
\begin{array}{ccl}
\dot{p} &=& -\displaystyle \frac{\partial H}{\partial q}  - g \left( \beta D g \frac{\partial H}{\partial p} - D g' -  \hat{R} \right), \\
\dot{q} &=& \displaystyle \frac{\partial H}{\partial p},
\end{array}
\right.
\label{eq_generalized_Langevin}
\end{equation}
which always satisfies the H-theorem for an arbitrary function $g(p)$.
Note that the equation of motion~(\ref{eq_generalized_Langevin})
is not invariant for time-reversal operation, since the friction term $h(p)$ changes
its sign for time-reversal operation, while $\dot{p}$ does not. 
Equation~(\ref{eq_generalized_Langevin}) is, therefore, always time-irreversible.
The Fokker-Planck equation corresponding to Eq.~(\ref{eq_generalized_Langevin}) is 
given by
\begin{equation}
\frac{\partial f}{\partial t} = - \{ H, f\}
+ \frac{\partial}{\partial p} \left[ D g^2
\left(
\beta \frac{\partial H}{\partial p}f + \frac{\partial f}{\partial p}
\right)
\right]
.
\label{eq_Fokker_Planck}
\end{equation}
One can confirm that the Gibbs canonical distribution $f \propto \exp(-\beta H)$ can be achieved when 
the system reaches the steady state, {\it i.e.}, $\partial f/\partial t = 0$.

Next, we consider the relationship between the H-theorem and a
thermostated system with coarse graining.
Consider the system described by the Hamiltonian $H(p,q)$ with the Nos\'e-Hoover
thermostat.
The equations of motion of this system are
\begin{equation}
\left\{
\begin{array}{ccl}
\dot{p} &=& -\displaystyle \frac{\partial H}{\partial q}  - p\zeta, \\
\dot{q} &=& \displaystyle \frac{\partial H}{\partial p}, \\
\dot{\zeta} &=&  p  \displaystyle \frac{\partial H}{\partial p} - \frac{1}{\beta},
\end{array}
\right.
\end{equation}
with an additional degree of freedom $\zeta$.
Note that the relaxation time of the thermostat is set to be unity
and the general form of Hamiltonian is considered.\cite{Hamiltonian}
These equations of motion satisfy the continuum equation,
\begin{equation}
- \frac{\partial }{\partial p} \left(\dot{p} f\right)
- \frac{\partial }{\partial q} \left(\dot{q} f\right)
- \frac{\partial }{\partial \zeta} \left(\dot{\zeta} f\right) = 0, 
\end{equation}
for the canonical distribution
\begin{equation}
f(p,q,\zeta) = Z^{-1} \exp(-\beta H_{\mathrm ex}),
\end{equation}
with the extended Hamiltonian $H_{\mathrm ex} \equiv H + \zeta^2/2$.
Therefore, the Nos\'e--Hoover equation keeps the canonical distribution invariant.

We can eliminate $\zeta$ formally as
\begin{eqnarray}
\left\{
\begin{array}{lcl}
\dot{p} &=& - \displaystyle \frac{\partial H}{\partial q} - p \int_0^t G(s)  \diff s, \\
\dot{q} &=& \displaystyle \frac{\partial H}{\partial p},
\end{array}
\right. \label{eq_nh}
\end{eqnarray}
where
\begin{equation}
G(t) \equiv p \frac{\partial H}{\partial p} - \frac{1}{\beta}.
\end{equation}
Equation~(\ref{eq_nh}) has the form of the Langevin equation with the friction term exhibiting a memory effect.
The time evolution is time-reversible and non-Markovian because of this memory effect.
In the following, we shall derive a stochastic--Nos\'e--Hoover equation by 
applying the Markovian approximation to the friction term.
$G(t)$ varies more slowly than $p$ in the many-body system
and should fluctuate around zero, since
\begin{equation}
\left< p \frac{\partial H}{\partial p} \right> = \frac{1}{\beta},
\end{equation}
where angular brackets denote the average over the canonical ensemble.
Assuming that $G(t)$ is not correlated to time longer than $\tau$,
we can approximate the integration in Eq.~(\ref{eq_nh}) as
\begin{equation}
\int_0^t G(s) \diff s \sim \tau G(t) + \hat{R}(t), \label{eq_markovian_approximation}
\end{equation}
with the Gaussian noise $\hat{R}$ defined in Eq.~(\ref{eq_noise}).
Inserting Eq.~(\ref{eq_markovian_approximation}) to Eq.~(\ref{eq_nh}),
we obtain the stochastic--Nos\'e--Hoover equation as
\begin{equation}
\left\{
\begin{array}{lll}
\dot{p} &=& - \displaystyle \frac{\partial H}{\partial q} - p
\left( \tau p \frac{\partial H}{\partial p} - \frac{\tau}{\beta} +\hat{R}\right).\\
\dot{q} &=& \displaystyle \frac{\partial H}{\partial p}.\\
\end{array}
\right. \label{eq_stochastic_nh}
\end{equation}
The characteristic time scale $\tau$ should be a
function of the inverse temperature $\beta$ and noise amplitude $D$,
and therefore, we can assume the Einstein relation $\tau = 1/(\beta D)$.
Thus, the Fokker-Planck equation corresponding to Eq.~(\ref{eq_stochastic_nh}) is given by
\begin{equation}
\frac{\partial f}{\partial t} = - \{ H, f\}
+ \frac{\partial}{\partial p} \left[ \frac{p^2}{\beta^2 D}
\left(
\beta \frac{\partial H}{\partial p}f + \frac{\partial f}{\partial p}
\right)
\right]
.
\end{equation}
This equation is the special case of Eq.~(\ref{eq_Fokker_Planck}) with $g(p) = p/(\beta D)$;
therefore, the free energy of this system decays monotonically and
the canonical distribution $f \sim \exp(-\beta H)$ is achieved 
when the system reaches the steady state.

The elimination of the additional variable $\zeta$
can be considered as the projection from the phase space $(p, q, \zeta)$
to its subspace $(p, q)$.
Equation (\ref{eq_nh}) corresponds to the Mori equation that is time-reversible
and contains the memory effect.
The Markovian approximation breaks time reversibility explicitly, as shown in Eq.~(\ref{eq_markovian_approximation}),
since the left-hand side changes its sign for time-reversal operation,
while the right-hand side does not.
This shows how the thermostated system loses its time reversibility 
with the Markovian approximation. 
It is worth noting that the Markovian approximation
applied in Eq.~(\ref{eq_markovian_approximation}) is justified only 
in the system with many degrees of freedom, while we have shown 
the derivation of a single degree of freedom.

To summarize, a generalized Langevin equation with
state-dependent random force is derived when the H-theorem holds in the system.
The Nos\'e--Hoover method coarse-grained with time satisfies the H-theorem,
and therefore, works as a heat bath.

We have studied the multiplicative noise in order to consider the
generalized Langevin equation that includes the stochastic--Nos\'e--Hoover
equation. Recently, a Langevin equation with a random multiplicative noise
has been studied~\cite{Sakaguchi2001}. This study shows that the process exhibits
Tsallis statistics and the H-theorem is satisfied in this system.
Moreover, a deterministic thermostat,
which achieves a Tsallis distribution as a result of time evolution, has been proposed~\cite{TsallisDynamics}.
Therefore, it is interesting to note that this is one of the issues
that must be further investigated to study the entropy production of a
system with general thermostats in a 
manner similar to that presented in this article.

The author would like to thank N. Ito, K. Hayashi and H. Kobayashi for fruitful discussions.
The present study was supported by the 21st COE program,
``Frontiers of Computational Science," Nagoya University, and by KAKENHI (19740235) and (19540400).


\begin{thebibliography}{99}
\bibitem{Hamiltonian} When the Hamiltonian of the natural form $H = p^2/(2m) + V(q)$ is assumed,
the form of $\dot{\zeta}$ is typically reduced to be $\dot{\zeta} = p^2/m - 1/\beta$.
\end{thebibliography}
\end{document}